\documentclass[11pt,twoside]{article}

\usepackage{asp2006}
\usepackage{epsf}
\usepackage{graphicx}
\usepackage{lscape}
\usepackage{natbib}

\begin{document}
\title{EUCLID : Dark Universe Probe and Microlensing planet Hunter}   
\author{J.P. Beaulieu\altaffilmark{1,2,3},D.P. Bennett\altaffilmark{4,3}, V Batista\altaffilmark{1,3}, A Cassan\altaffilmark{1,3}, D. Kubas\altaffilmark{1,3},
P. Fouqu\'e\altaffilmark{5,3}, E. Kerrins\altaffilmark{6}, S. Mao\altaffilmark{6}, J. Miralda-Escud\'e\altaffilmark{7}, J. Wambsganss\altaffilmark{8}, B.S. Gaudi\altaffilmark{9}, A. Gould\altaffilmark{9,3} and S. Dong\altaffilmark{10} }    
\altaffiltext{1} {Institut d'Astrophysique de Paris, 75014 Paris, France}  
\altaffiltext{2} {Department of Physics and Astronomy, University College London, London WC1E 6BT, UK}   \altaffiltext{3} {HOLMES collaboration}    
\altaffiltext{4}{University of Notre Dame, Department of Physics, Notre Dame IN 46556, USA}    
\altaffiltext{5}{Observatoire Midi-Pyr\'en\'ees, UMR 5572, 31400 Toulouse, France}    
\altaffiltext{6}{Jordrell Bank Center for Astrophysics, Univ of Manchester, Manchester, UK}    
\altaffiltext{7}{ICREA/ICC-IEEC, Univ of Barcelona, Barcelona, Spain}    
\altaffiltext{8}{Astronomisches Rechen-Institut, Zentrum fur Astronomie, 69120 Heidelberg, Germany}    
\altaffiltext{9}{Department of Astronomy, Ohio State University, Columbus OH 43210, USA}    
\altaffiltext{10}{Institute for advance studies, School of Natural Sciences, Princeton NJ 08540, USA}    

\begin{abstract} 
There is a remarkable synergy between requirements for Dark Energy probes by cosmic shear measurements and planet hunting by microlensing. Employing weak and strong gravitational lensing to trace and detect the distribution of matter on cosmic and Galactic scales,  but as well as to the very small scales of exoplanets is a unique meeting point from cosmology to exoplanets.  It will use gravity as the tool to explore the full range of masses not accessible by any other means.
EUCLID is a 1.2m telescope with  optical and IR wide field imagers and slitless spectroscopy, proposed to ESA Cosmic Vision to probe for Dark Energy, Baryonic acoustic oscillation, galaxy evolution, and an exoplanet hunt via microlensing.
A 3 months microlensing program will already efficiently probe for planets down to the mass of Mars
at the snow line, for free floating terrestrial or gaseous planets and habitable super Earth. 
A 12+ months survey would give a census on habitable Earth planets around solar
like stars.  This is the perfect complement to the statistics that will be provided by the KEPLER satellite, 
and these missions combined will provide a full census of extrasolar planets from hot, warm, habitable, frozen to free floating.
\end{abstract}



\section{Introduction}



In the last fifteen years, astronomers have found over 400 exoplanets (Schneider 2009), including some in systems that resemble our very own solar system (Gaudi et al., 2008). These discoveries have already challenged and revolutionized our theories of planet formation and dynamical evolution. 
Several different methods have been used to discover exoplanets, including radial velocity, stellar transits, and gravitational microlensing. Exoplanet detection via gravitational microlensing is a relatively new method (Mao and Paczynski, 1991, Gould and Loeb, 1992, Wambsganss, 1997) and is based on Einstein’s theory of general relativity. So far 9 exoplanets have been published with this method. While this number is relatively modest compared with that discovered by the radial velocity method, microlensing probes a part of the parameter space (host separation vs. planet mass) not accessible in the medium term to other methods (see Figure 1.). 

The mass distribution of microlensing exoplanets has already revealed that  cold super-Earths (at or beyond the “snow line” and with a mass of around 5 to 15$M{_\oplus}$) appear to be common (Beaulieu et al., 2006, Gould et al., 2006, Gould et al., 2007, 
Kubas et al., 2008, Bennett 2010, this volume). Microlensing is currently capable of detecting cool planets of super-Earth mass from the ground and, with a network of wide-field telescopes strategically located around the world, could detect planets with mass as low as the Earth. Old, free-floating planets can also be detected;  a significant population of such planets are expected to be ejected during the formation of planetary systems (Juric and Tremaine, 2008).
Microlensing is roughly uniformly sensitive to planets orbiting all types of stars, as well as white dwarfs, neutron stars, and black holes, while other method are most sensitive to FGK dwarfs and are now extending to M dwarfs. 
It is therefore an independent and complementary detection method for aiding a comprehensive understanding of the planet formation process. Ground-based microlensing mostly probes exoplanets outside the snow line, where the favoured core accretion theory of planet formation predicts a larger number of low-mass exoplanets (Ida and Lin, 2005). The statistics provided by microlensing will enable a critical test of the core accretion model. 

Exoplanets probed by microlensing are much further away than those probed with other methods. They provide an interesting comparison sample with  nearby exoplanets, and allow us to study the extrasolar
population throughout the Galaxy. In particular, the host stars with exoplanets appear to have higher metallicity (e.g. Fischer and Valenti, 2005). Since the metallicity is on average higher as one goes towards the Galactic centre, the abundance of exoplanets may well be somewhat higher in microlensing surveys.

\section{ Basic microlensing principles}
The physical basis of microlensing is the deflection of light rays by a massive body.  A distant source star is temporarily magnified by the gravitational potential of an intervening star (the lens) passing near the line of sight, with an impact parameter smaller than the Einstein ring radius $R_E$, a quantity which depends on the mass of the lens, and the geometry of the alignment. For a source star in the Bulge, with a 0.3 $M_\odot$ lens, $R_E$ $\sim$ 2 AU, the angular Einstein ring radius is $\sim$1 mas, and the time to transit $R_E$ is typically 20-30 days, but can be in the range 5-100 days. 
The lensing magnification is determined by the degree of alignment of  the lens and source stars.   The closer the alignment the higher the magnification.

\begin{figure}[!ht]
\includegraphics[width=13 truecm]{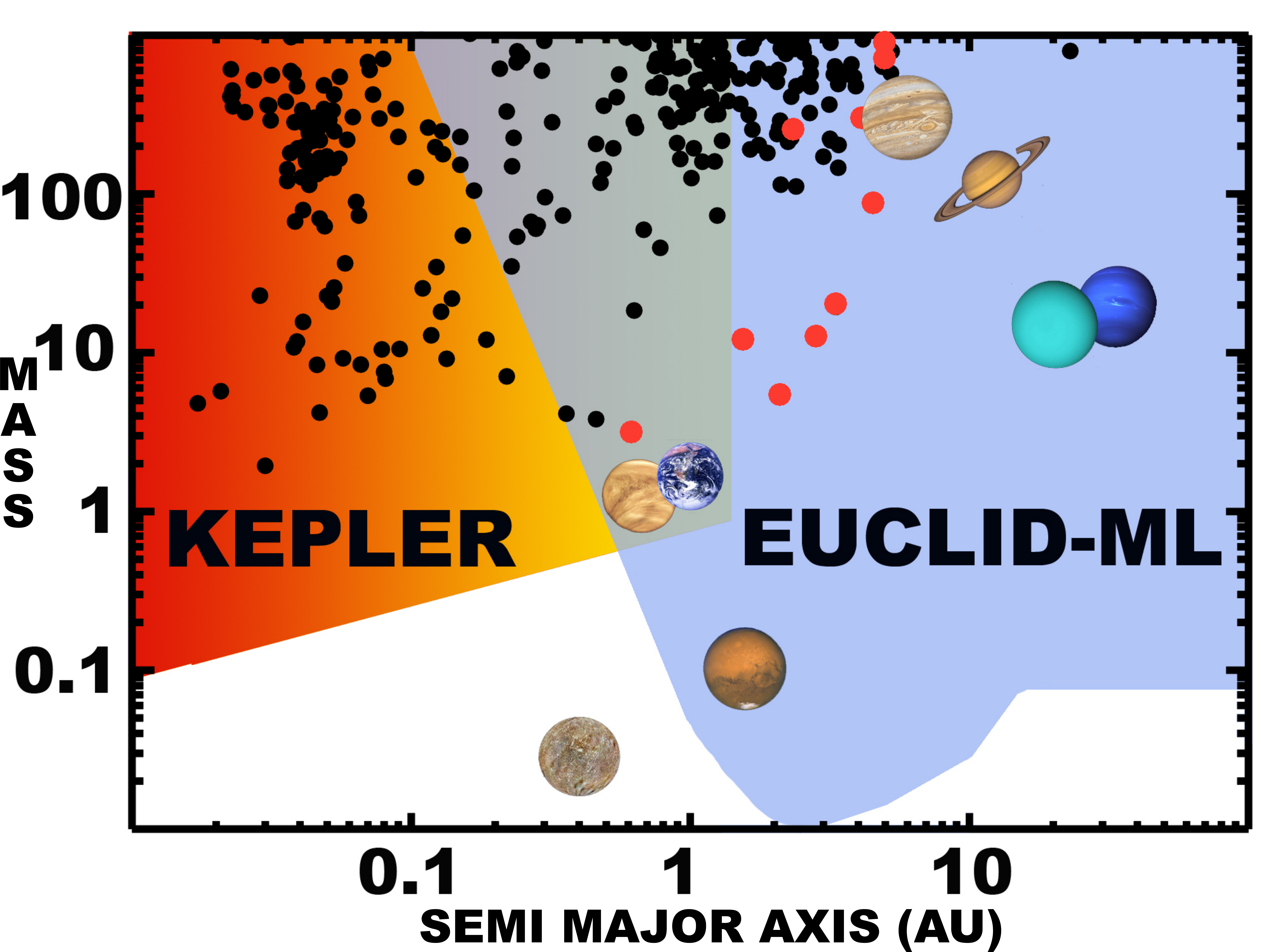}
\caption{Semi major axis as a function of mass for all exoplanets discovered as of September 2009 (microlensing planets
are plotted as red dots) and the planets from our solar system. We also plot the sensitivity of KEPLER and of space based microlensing 
observations.}
\label{param}
\end{figure}

A planetary companion to the lens star will induce a perturbation to the microlensing light curve with a duration that scales  with the square root of the planet’s mass, lasting typically a few hours (for an Earth) to a few days (for a Jupiter). Hence, planets can be discovered by dense photometric sampling of ongoing microlensing events (Mao and Paczynski 1991, Gould and Loeb 1992). The limiting mass for the microlensing method occurs when the planetary Einstein radius becomes smaller than the projected radius of the source star (Bennett and Rhie1996). The $\sim 5.5 M_\oplus$ planet detected by Beaulieu et al., (2006) 
is near this limit for a giant source star, but most microlensing events have G or K-dwarf source stars with radii that are at least 10 times smaller than this. 
High angular enough resolution to resolve dwarf sources of the galactic bulge ($\leq 0.5$ arcsec) will open the sensitivity below a few Earth masses (Figure 2).

\begin{figure}
\includegraphics[width=6.5 truecm]{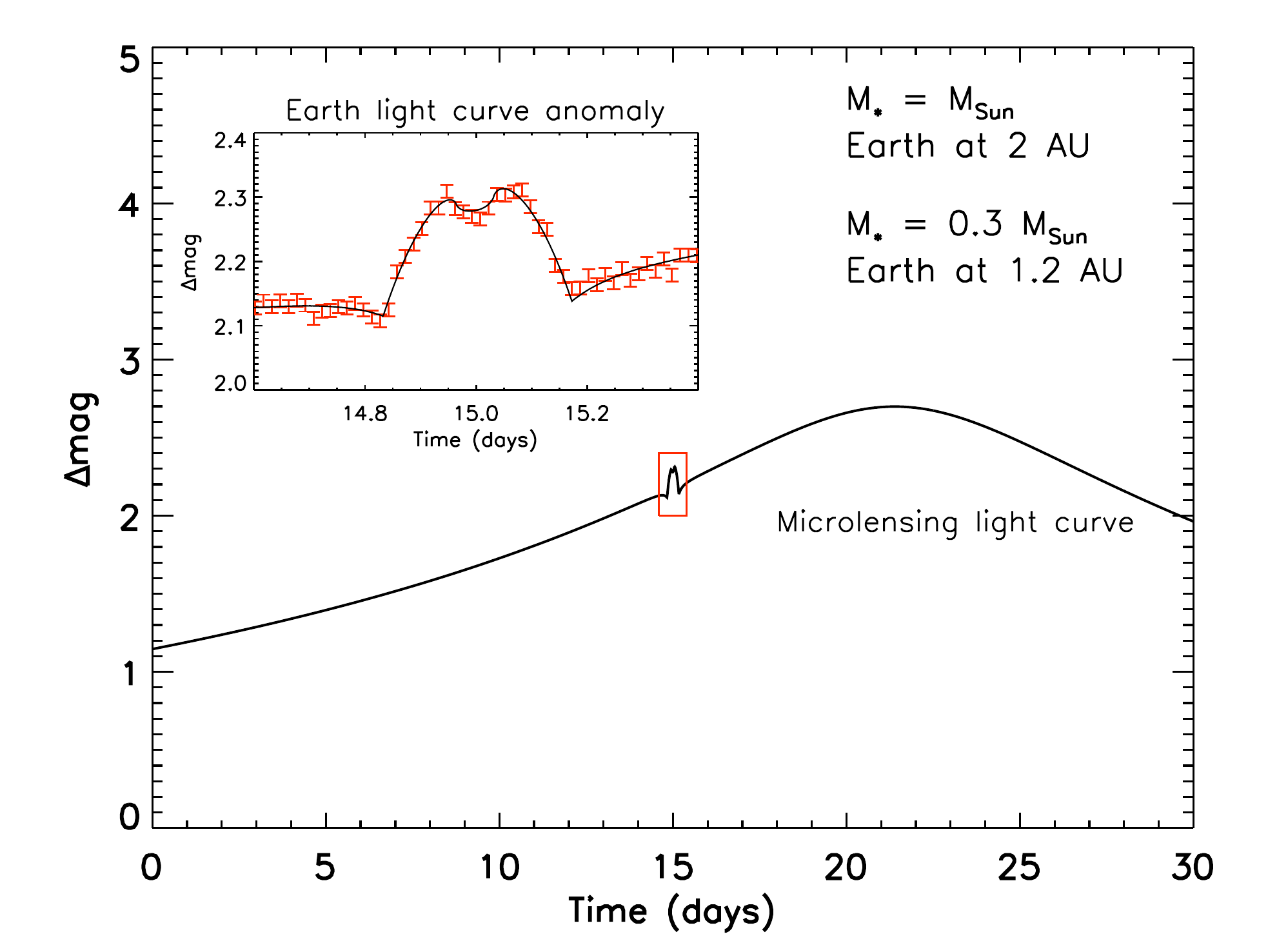}\includegraphics[width=6.5 truecm]{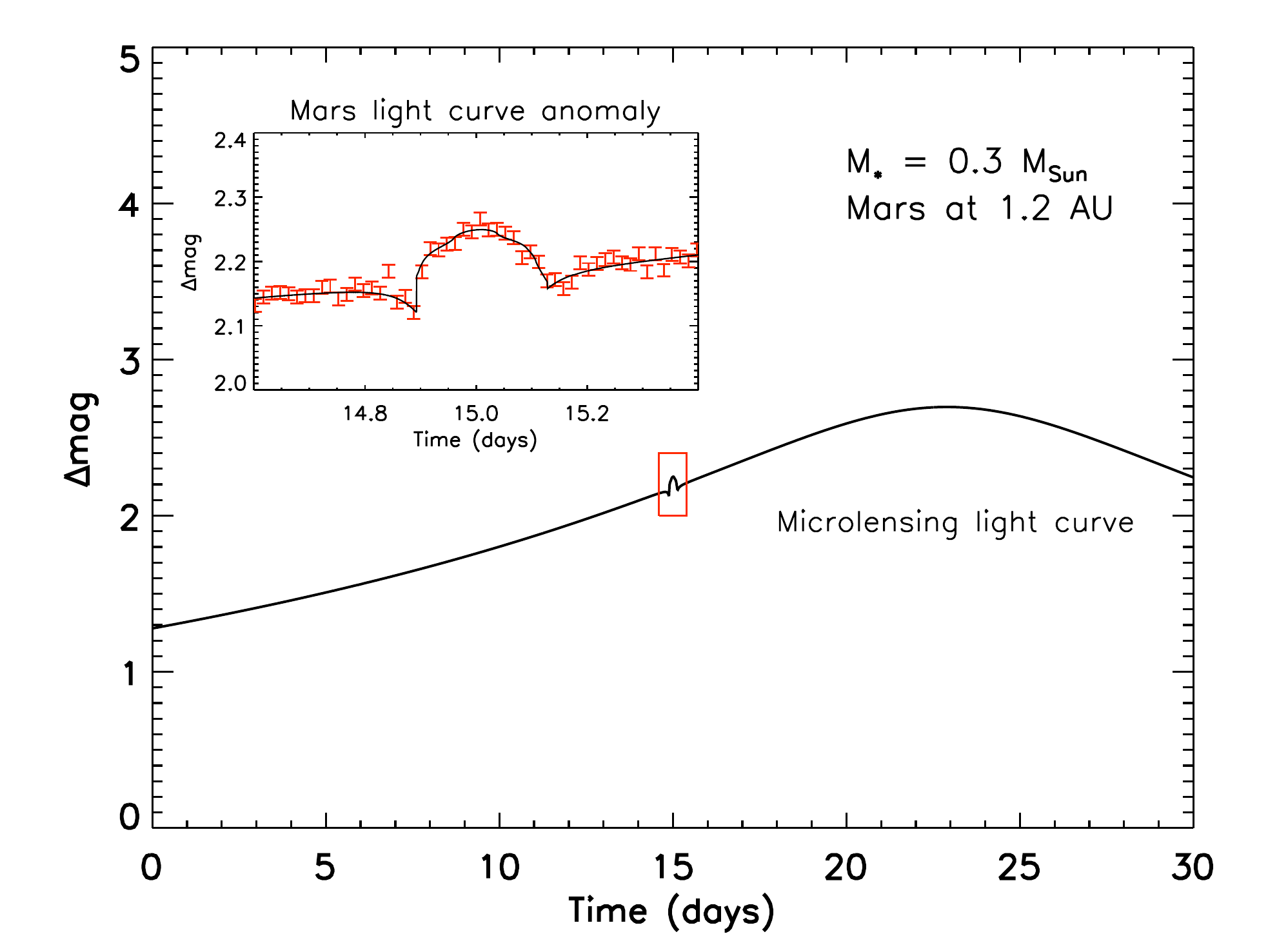}
\caption{The two figures illustrate the detection capability of the
  microlensing technique in the very low-mass exoplanet regime. Here, the
  source star and the lens (the planet host star) are both located in the
  Galactic Bulge. The sampling interval is twenty minutes, and the
  photometric precision is one percent. The planetary signal on the
  left figure is expected from an Earth-mass planet at 2 AU around a
  solar star, or from an Earth at 1.2 AU but orbiting an
  $0.3\,M_\odot$ M-dwarf star. 
  A planet of the mass of Mars ($0.1 M_\oplus$) at 1.2 AU can also be
  detected around such a low-mass host star (right figure). These are typical examples of low mass telluric planets to be detected by EUCLID.}
\label{EARTH-MARS}
\end{figure}

The inverse problem, finding the properties of the lensing system (planet/star mass ratio, star-planet projected separation) from an observed light curve, is a complex non-linear one within a wide parameter space. In general, model distributions for the spatial mass density of the Milky Way, the velocities of potential lens and source stars, and a mass function of the lens stars are required in order to derive probability distributions for the masses of the planet and the lens star, their distance, as well as the orbital radius and period of the planet by means of Bayesian analysis. With complementary high angular resolution observations, currently done either by HST or with adaptive optics, it is possible to get additional constraints to the parameters of the system, and determine masses to 10 \% by directly constraining the light coming from the lens and measuring the lens and source relative proper motion (Bennett et al. 2006, Bennett et al.,  2007,Dong et al., 2009). 
A space-based microlensing survey can provide the planet mass, projected separation from the host, host star mass and its distance from the observer for most events using this method.\\

Different papers have presented the future strategies in the near, medium  and long term, with the ultimate goal of achieving a full census of Earth-like planets with either a dedicated space mission (Microlensing Planet Finder, MPF) or advocating for synergy between Dark Energy Probes and microlensing. There is a general consensus in the microlensing community about these mile stones,
and this consensus has been endorsed by the US ExoPlanet Task Force (ExoPTF)".
 White papers submitted to the ExoPTF (Bennett et al., 2007,Gould et al., 2007),
the exoplanet forum (Gaudi et al., 2009a), the JDEM request for 
information, ESA-EPRAT (ExoPlanetary Roadmap Advisory Team) (Beaulieu et al., 2008) and 
Astro2010 PSF (Bennett et al. 2009,Gaudi et al., 2009b), and the Pathways conference in Barcelona (Bennett 2010). 

\section{A program on board EUCLID to hunt for planets}

\noindent {\bf Space based microlensing observations} \\
The ideal satellite is a 1m class space telescope with a focal plane of $0.5$ square degree
 or more in the visible or in the near infra red. The Microlensing Planet Finder 
is an example of such a mission (Bennett et al. 2007), which has been proposed to NASA’s Discovery program, and endorsed by the ExoPTF.
Despite the fact that the designs were completely independent, there is a remarkable similarity between the requirements for missions aimed at probing Dark Energy via cosmic shear (Refregier et al., 2010) and a microlensing planet hunting mission (Beaulieu et al., 2008, Bennett et al. 2007, 2009). 
EUCLID is a proposed mission to measure
parameters of dark energy using weak gravitational lensing and baryonic acoustic
oscillation, test the general relativity and the Cold Dark Matter paradigm for
structure formation submitted to the ESA COSMIC VISION program. It is a
1.2m Korsch telescope with in particular a 0.48 square degree imager in a broad
optical band consisting of R+I+Z (0.1 arcsec per pixel) and in the Y, J, H band
(0.3 arcsec per pixel). Microlensing benefits from the strong requirement from
cosmic shear on the imaging channel, and does not add any constraint to the
design of EUCLID.

\noindent {\bf Observing strategy} \\
We will monitor 2 square degree of the area with highest optical depth to microlensing from the galactic Bulge with a sampling rate once every twenty  minutes. Observations will be conducted in the optical and NIR channel.

\noindent {\bf Angular resolution is the key to extend sensitivity below few earth masses} \\
Microlensing relies upon the high density of source and lens stars towards the Galactic bulge to generate the stellar alignments that are needed to
generate microlensing events, but this high star density also means that the bulge main sequence
source stars are not generally resolved in groundbased images. This means
that the precise photometry needed to detect planets of $\leq 1M_\oplus$ is not possible from the ground unless the
magnification due to the stellar lens is moderately high. This, in turn, implies that ground-based
microlensing is only sensitive to terrestrial planets located close to the Einstein ring (at $\sim$2-3 AU). The
full sensitivity to terrestrial planets in all orbits from $0.5 AU$ to free floating comes only from a space-based survey (Figure 1).
In figure 2 we give examples of simulated detections of an Earth and a Mars-mass planet.

\noindent {\bf Microlensing from space yields precise star and planet parameters} \\
The high angular resolution and stable point-spread-functions available
from space enable a space-based microlensing survey to detect most of
the planetary host stars. When combined with the microlensing light
curve data, this allows a precise determination of the planet and star
properties for most events (Bennett et al. 2007).

{\bf Probing a parameter space out of reach of any other technique} \\
The Exoplanet Task Force (ExoPTF) recently released a report (Lunine et al., 2008) that
evaluated all of the current and proposed methods to find and study exoplanets, and they
expressed strong support for space-based microlensing. Their finding regarding space-based
microlensing states that: {\sl “Space-based microlensing is the optimal approach to providing a true
statistical census of planetary systems in the Galaxy, over a range of likely semi-major axes, and
can likely be conducted with a Discovery-class mission.”} It can also be accomplished as
a program on board the EUCLID M class mission with Dark Energy probe as primary objective.

A EUCLID microlensing survey provides a census of extrasolar planets that is complete (in
a statistical sense) down to $0.1M_\oplus$ at orbital separations $\geq$ 0.5 AU. When combined with the
results of the Kepler mission (and ground based radial velocity surveys)
EUCLID  will give a comprehensive
picture of all types of extrasolar planets with masses down to well below an Earth mass.
 This fundamental exoplanet census data is needed to gain a
comprehensive understanding of processes of planet formation and migration, and this
understanding of planet formation is an important ingredient for the understanding of the
requirements for habitable planets and the development of life on extrasolar planets.

A subset of the science goals can be accomplished with an enhanced ground-based
microlensing program (Gaudi et al. 2009ab), which would be sensitive to Earth-mass planets in the
vicinity of the “snow-line”. But such a survey would have its sensitivity to Earth-like planets
limited to a narrow range of semi-major axes, so it would not provide the complete picture of the
frequency of exoplanets down to $0.1 M_\oplus$ that a space-based microlensing survey would provide.
Furthermore, a ground-based survey would not be able to detect the planetary host stars for most
of the events, and so it will not provide the systematic data on the variation of exoplanet
properties as a function of host star type that a space-based survey will provide.

{\bf Duration of the program} \\

One of the remarkable feature of the EUCLID microlensing program is its linear sensitivity to allocated 
time and area of the focal plane. The minimal time allocation of three months will already give important
statistics on planets at the snow line, down to the mass of mars, and of free floating planets. 
Habitable super Earth will also be probed. Longer observing time (12 months of galactic bulge observing) would lead to sensitivity to a true analogue habitable Earth mass planets orbiting solar like stars.


{\bf Reference} \\
         Beaulieu et al., 2006, Nature 439, 437 \\
         Beaulieu et al., 2008, ESA EPRAT, astroph 0808.0005 \\
        Bennett, D.~P.,Anderson, J. and Gaudi, B.~S., 2007, ApJ 660, 781 \\
       Bennett, D.P., et al., 2007, astroph 0704.0454 \\
   Bennett D.P., et al., 2009, astroph 0902.3000 \\
   Dong S., et al., 2009, ApJ 695, 970 \\
  Fischer D.~A. and Valenti, J., 2005, ApJ 622, 1102 \\
  Gaudi B.S. et al. 2008, Science 319, 927 \\
   Gaudi B.S. et al. 2009a, in "{Exoplanet Community Report on Microlensing}"\\
   Gaudi B.S., et al., 2009b, astroph 0903.0880 \\
   Gould A., \& Loeb A., 1992, ApJ 396, 104 \\
   Gould A., et al., 2006, ApJ 644, L37 \\
   Gould A., Gaudi B.S., Bennett D.P., 2007, astroph 0704.0767 \\
    Ida S. and Lin D.N.C., 2005, Apj 626, 1045 \\
   Juric M. and Tremaine S., 2008, ApJ 686, 603 \\
   Kubas D., et al., 2008, AA 483, 317 \\
   Lunine J., et al., 2008, in  "{Exoplanet Task Force Report}'', astroph 0808.2754 \\
    Mao S. and Paczynski B., 1991, ApJ 374, L37 \\
    Refregier A., et al. 2010, astroph 1001.0061 \\
   Schneider J., 2009, extrasolar planet encyclopedia \\
   Wambsganss J., 1997, MNRAS 284, 172 \\

\end{document}